\documentclass[prd,preprintnumbers,floatfix,aps,notitlepage,nofootinbib,twocolumn,showpacs,amssymb]{revtex4-2}
\usepackage{amsmath,amsfonts,bm}
\usepackage{graphicx}
\usepackage{cancel}
\usepackage[colorlinks, linkcolor={red},citecolor={blue}]{hyperref}
\usepackage{xcolor}

\newcommand{\re}{r_{\rm e}} 
\begin{document}
\title{Regular Schwarzschild black holes and cosmological models}
\author{Roberto Casadio}
\email{casadio@bo.infn.it}
\affiliation{Dipartimento di Fisica e Astronomia,
Alma Mater Universit\`a di Bologna,
40126 Bologna, Italy
\\
Istituto Nazionale di Fisica Nucleare, I.S.~FLaG
Sezione di Bologna, 40127 Bologna, Italy
\\
Alma Mater Research Center on Applied Mathematics (AM$^2$),
Via Saragozza 8, 40123 Bologna, Italy}
\author{Alexander Kamenshchik}
\email{kamenshchik@bo.infn.it}
\affiliation{Dipartimento di Fisica e Astronomia,
Alma Mater Universit\`a di Bologna,
40126 Bologna, Italy
\\
Istituto Nazionale di Fisica Nucleare, I.S.~FLaG
Sezione di Bologna, 40127 Bologna, Italy}
\author{Jorge Ovalle}
\email[]{jorge.ovalle@physics.slu.cz}
\affiliation{Research Centre for Theoretical Physics and Astrophysics,
Institute of Physics, Silesian University in Opava, CZ-746 01 Opava,
Czech Republic.}
\begin{abstract}
We study regular Schwarzschild black holes in General Relativity as an alternative
to the singular counterpart.
We analyze two types of solutions which are completely parameterised by the ADM
mass alone. 
{We find that both families of regular solutions}
contain a de Sitter condensate at the core and admit (quasi) extremal
black hole configurations in which the two horizons are arbitrarily close.
Cosmological models based on these regular configurations are also analyzed,
finding that they describe non-trivial Kantowski-Sachs universes
{ free of singularities}.
\end{abstract} 
\maketitle
%
%
%
\section{Introduction}
\label{sec:intro}
It is well known that general relativity (GR) predicts singularities inside trapped
regions, as stated in the singularity theorems~\cite{Penrose:1964wq,Hawking:1973uf},
based on general and quite reasonable assumptions about the matter source.
If we accept the weak cosmic censorship conjecture~\cite{Penrose:1969pc} to
exclude the presence of naked singularities in nature, this means that the end
of the collapse could well be described by vacuum black hole (BH) solutions of GR.
\par
On the other hand, one can consider types of matter that circumvent the
singularity theorems, allowing the collapse to form an event horizon without
leading to a singularity.
This is the case of regular (that is completely non-singular) BHs.
Unfortunately, this inevitably leads to the existence of at least a second horizon,
the so-called Cauchy horizon, which has proven particularly
problematic~\cite{Poisson:1989zz,Poisson:1990eh} (see also Refs.~\cite{Ori:1991zz,
Carballo-Rubio:2018pmi,Bonanno:2020fgp,Carballo-Rubio:2021bpr,Carballo-Rubio:2022kad,
Franzin:2022wai,Casadio:2022ndh,Bonanno:2022jjp,Casadio:2023iqt,Ovalle:2023vvu} for recent studies),
thus motivating the strong cosmic censorship conjecture~\cite{Penrose:1969pc}.
\par
If we ignore for now all the problems associated with the Cauchy horizon and focus
mainly on the construction and analysis of regular BHs, we will quickly realize that
their production remains relatively easy (by employing still reasonable forms of matter).
It is even possible to describe them in terms of a nonlinear electrodynamic
theory~\cite{Ayon-Beato:1998hmi}.
Unfortunately, the above does not shed much light on the fundamental problem, 
{\em i.e.}~the details of how the singularity is formed and, above all, how it is avoided.
These are critical aspects in order to investigate the validity of GR under conditions
of extreme curvature.
If these non-singular configurations really form during the collapse, effects associated
with the Cauchy horizon seem to indicate that they are unstable, and would therefore
represent at best a transitory state in the eventual formation of a singularity. 
\par
All the questions mentioned above require a detailed study of the inner BH region.
Most cases exhibit a simple internal geometry, which 
is of course consistent with the ultimate state of the collapse.
However, as argued in Ref.~\cite{Ovalle:2024wtv}, the interior need
not have this extreme simplicity, since the weak cosmic censorship conjecture
establishes the formation of the event horizon before the singularity appears,
thus allowing for more complex internal structures than the eventual final singularity.
This is precisely the case reported in detail in Ref.~\cite{Ovalle:2024wtv},
where alternative sources for the exterior region of the Schwarzschild BH in GR
were investigated.
Among the most attractive characteristics of these solutions, we highlight:
(i) they depend on one parameter, namely, the total ADM mass ${\cal M}$
(no primary hair);
(ii) no form of exotic matter is present;
(iii) the space-time is continuous across the horizon without additional
structures on it (like a thin shell);
(iv) tidal forces are finite everywhere for (integrable) singular
solutions~\cite{Lukash:2013ts,Casadio:2023iqt};
(v) there are simple regular solutions which might be an alternative
to the Schwarzschild BH as the final stage of gravitational collapse. 
\par
{So far, only singular cases of revisited Schwarzschild BHs
have been considered, both in the cosmological context~\cite{Casadio:2024fol}
and for the analytical modelling of the gravitational collapse~\cite{Aoki:2024dyr}.
The aim of this work is therefore twofold:
first, to analyze in detail the case of regular Schwarzschild BHs and,
second, to study the cosmological models associated with these solutions.
The main difference is given by the unavoidable presence of an inner horizon
which introduces new features with respect to the singular cases previously
studied in Refs.~\cite{Casadio:2024fol,Aoki:2024dyr}.
About the first point, we will show that there exists (quasi) extremal configurations
with Schwarzschild exterior, despite the metrics only depend on one parameter ${\cal M}$.
For the second point, we are not aware of existing papers that exploit regular
BH solutions to generate cosmological models.
Again, the presence of two horizons makes the non-singular cases richer than
those corresponding to singular metrics that were studied in Ref.~\cite{Casadio:2024fol},
leading to (quasi) ciclic evolutions.}
\section{Inside the black hole}
\label{sec:insideBH}
We begin by reviewing briefly the approach employed in Ref.~\cite{Casadio:2024fol}
for static and spherically symmetric metrics of the Kerr-Schild form~\cite{kerrchild} 
\begin{equation}
ds^{2}
=
-f(r)\,dt^{2}
+\frac{dr^2}{f(r)}
+r^2\,d\Omega^2
\ ,
\label{metric}
\end{equation}
where
\begin{equation}
f
=
1-\frac{2\,m(r)}{r}
\ .
\label{gf}
\end{equation}
The Schwarzschild solution~\cite{Schwarzschild:1916uq} is obtained by setting
the Misner-Sharp-Hernandex mass function
\begin{equation}
m(r)
=
{\cal M}
\ ,
\quad
{\rm for}
\ r>0
\ ,
\label{mcond}
\end{equation}
where ${\cal M}$ is the ADM mass associated with a point-like singularity at the center $r=0$.
The coordinate singularity at $r=2\,{\cal M}\equiv\,h$ indicates the event horizon~\cite{Eddington:1924pmh,
Lemaitre:1933gd,Finkelstein:1958zz,Kruskal:1959vx,Szekeres:1960gm}.
\par
We slightly relax the condition~\eqref{mcond} to~\footnote{We shall denote $F(h)\equiv\,F(r)\big\rvert_{r=h}$
for any $F=F(r)$.
We shall also use units with $c=1$ and $\kappa=8\,\pi\,G_{\rm N}$.}
\begin{equation}
	\label{abcd}
m(r)
=
m(h)
=
{h}/{2}
=
{\cal M}
\ ,
\quad
{\rm for}\
r \geq h
\ ,
\end{equation}
so that the metric function~\eqref{gf} is given by
\begin{equation}
\label{mtransform}
f=\left\{
\begin{array}{l}
1-\frac{2\,m}{r}
\equiv f^-
\ ,
\quad
{\rm for}\
0< r \leq h
\\
\\
1-\frac{2\,{\cal M}}{r}
\equiv f^+
\ ,
\quad
{\rm for}\
r>h
\ .
\end{array}
\right.
\end{equation}
The system is governed by the Einstein-Hilbert action
\begin{equation}
\label{action}
S
=
\int\left(\frac{R}{2\,\kappa}+{\cal L}_{\rm M}\right)
\sqrt{-g}\,d^4x
\ ,
\end{equation}
with $R$ the scalar curvature and the Lagrangian density ${\cal L}_{\rm M}$
representing ordinary matter.
Eqs.~\eqref{mtransform} and~\eqref{action} imply that ${\cal L}_{\rm M}=0$
for $r>h$, and inside the horizon $0<r<h$ one finds the energy-momentum tensor 
\begin{eqnarray}
\label{emt}
T^\mu_{\ \nu}
=
{\rm diag}\left[p_r,-\epsilon,p_\theta,p_\theta\right]
\ ,
\end{eqnarray}
where the energy density $\epsilon$, radial pressure $p_r$ and transverse
pressure $p_\theta$ read~\footnote{Recall that the coordinates $t$ and $r$
exchange roles for $0<r<h$.}
\begin{eqnarray} 
\label{sources}
\epsilon=\frac{2\,{m}'}{\kappa\,r^2}
\ ,\quad
p_r=-\frac{2\,{m}'}{\kappa\,r^2}
\ ,\quad
p_\theta=-
\frac{{m}''}{\kappa\,r}
\ ,
\end{eqnarray}
where primes denote derivatives with respect to $r$.
Since Eqs.~\eqref{sources} are linear in the mass function ${m}$, 
any two solutions can be linearly combined, as a trivial case of
gravitational decoupling~\cite{Ovalle:2017fgl,Ovalle:2019qyi}. 
\par
From the contracted Bianchi identities $\nabla_\mu\,G^{\mu}_{\ \nu}=0$
one obtains the continuity equation
\begin{eqnarray}
\label{con111}
\epsilon'=
-\frac{2}{r}\left(p_\theta-p_r\right)
\ ,
\end{eqnarray}
which implies that $p_\theta>p_r$ if the energy density $\epsilon$ decreases
monotonically from the centre outwards ($\epsilon'<0$).
Continuity of the metric~\eqref{mtransform} across the horizon $r=h$
requires the matching conditions
\begin{equation}
\label{cond2}
m(h)={\cal M}
\ ,
\quad
m'(h)=0
\ ,
\end{equation}
and, from Eqs.~\eqref{sources} and~\eqref{cond2}, one must also have
\begin{equation}
\label{c2a}
\epsilon(h)=p_r(h)=0
\ .
\end{equation}
The tension $p_\theta$ can instead be discontinuous across $r=h$. 
\section{Regular black holes}
\label{sec:integrable}
In this section, we will analyze in detail the regular BHs with Schwarzschild
exterior, starting with the scalar curvature for the interior metric~\eqref{mtransform}, 
which reads
\begin{equation}
\label{R}
R
=
\frac{2\,r\,m''+4\,m'}{r^2}
\ ,
\quad
{\rm for}\
0< r\,\,{\leq}\,\,h
\ .
\end{equation}
In order to have a regular BH solution,
{we start by assuming~\cite{Ovalle:2024wtv}}
\begin{equation}
\label{R2}
R
=
\sum_{n=2}^{\infty}\,C_n\,r^{n-2}
\ ,
\quad
n\in\mathbb{N}
\ ,
\end{equation}
which, from Eq.~\eqref{R}, yields the mass function
\begin{equation}
\label{M}
m
=
M
-
\frac{Q^2}{2\,r}
+
\frac{1}{2}\,\sum_{n=2}^{\infty}\,\frac{C_n\,r^{n+1}}{(n+1)(n+2)}
\ ,
\end{equation}
for $0<r\leq\,h$, where $M$ and $Q$ are integration constants that can be identified
with the ADM mass of the Schwarzschild solution and a charge for the Reissner-Nordstr\"{o}m
(RN) geometry, respectively.
{In order to have a non-singular configuration, {\em i.e.},~the Ricci scalar~\eqref{R},
Ricci square $R_{\mu\nu}\,R^{\mu\nu}$ and Kretschmann scalar $R_{\mu\nu\rho\sigma}\,R^{\mu\nu\rho\sigma}$
that are regular around $r=0$, we must then impose}
\begin{eqnarray}
\label{Schw-limit2}
M=Q=0
\ .
\end{eqnarray}
\par
On the other hand, Eqs.~\eqref{sources} now yield
\begin{eqnarray}
\label{energy}
\kappa\,\epsilon
=
\sum_{n=2}^{\infty}\,\frac{C_n\,r^{n-2}}{n+2}
=
-\kappa\,p_r
\end{eqnarray}
and
\begin{eqnarray}
\kappa\,p_\theta
=
-\frac{1}{2}\sum_{n=2}^{\infty}\,\frac{n}{n+2}\,C_n\,r^{n-2}
\ ,
\label{pt}
\end{eqnarray}
for $0<r\leq\,h$. 
\subsection{Regular Schwarzschild BHs}
\label{SS:rBH}
\begin{figure}
\includegraphics[width=0.42\textwidth]{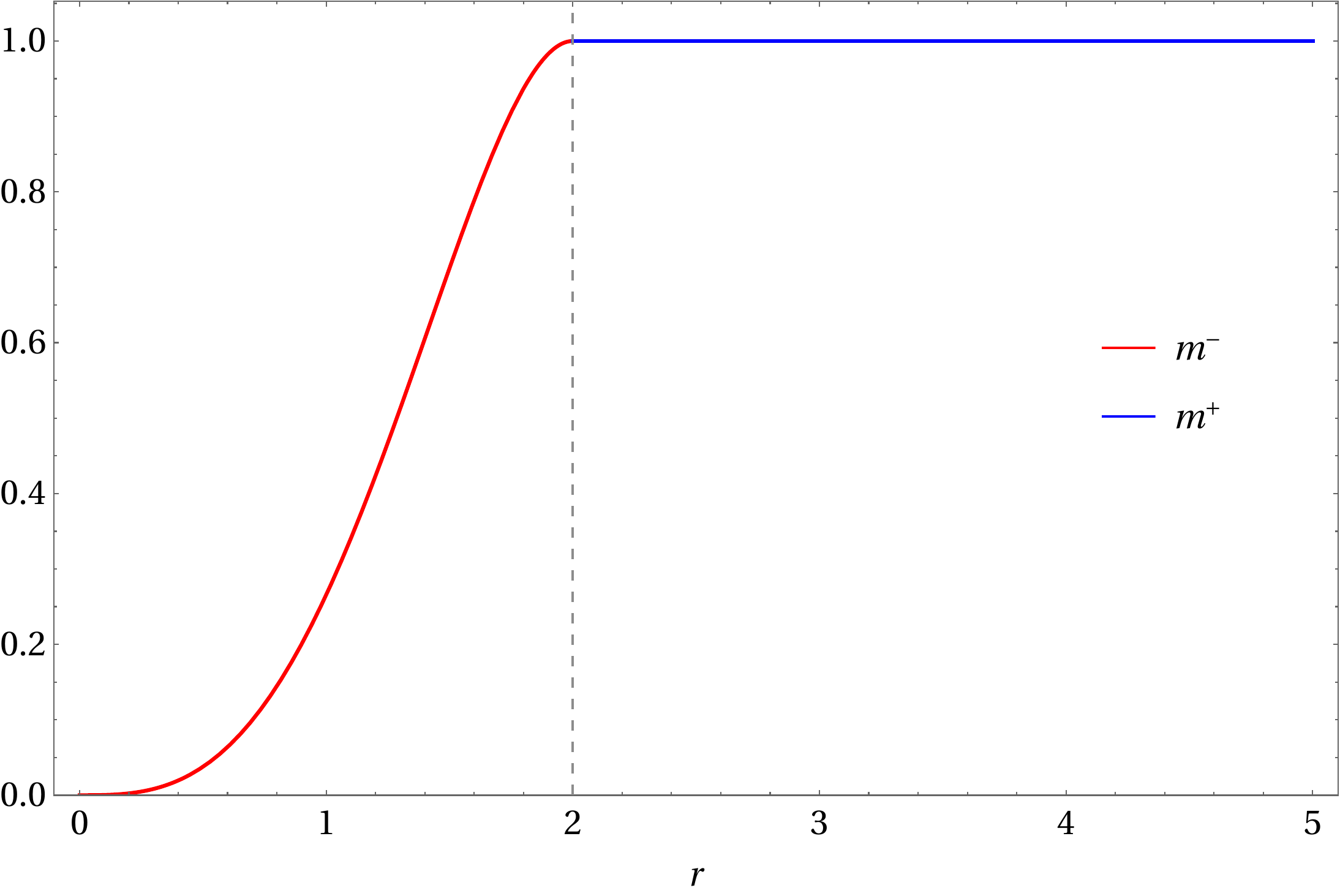}
\\
$\ $
\\
\includegraphics[width=0.42\textwidth]{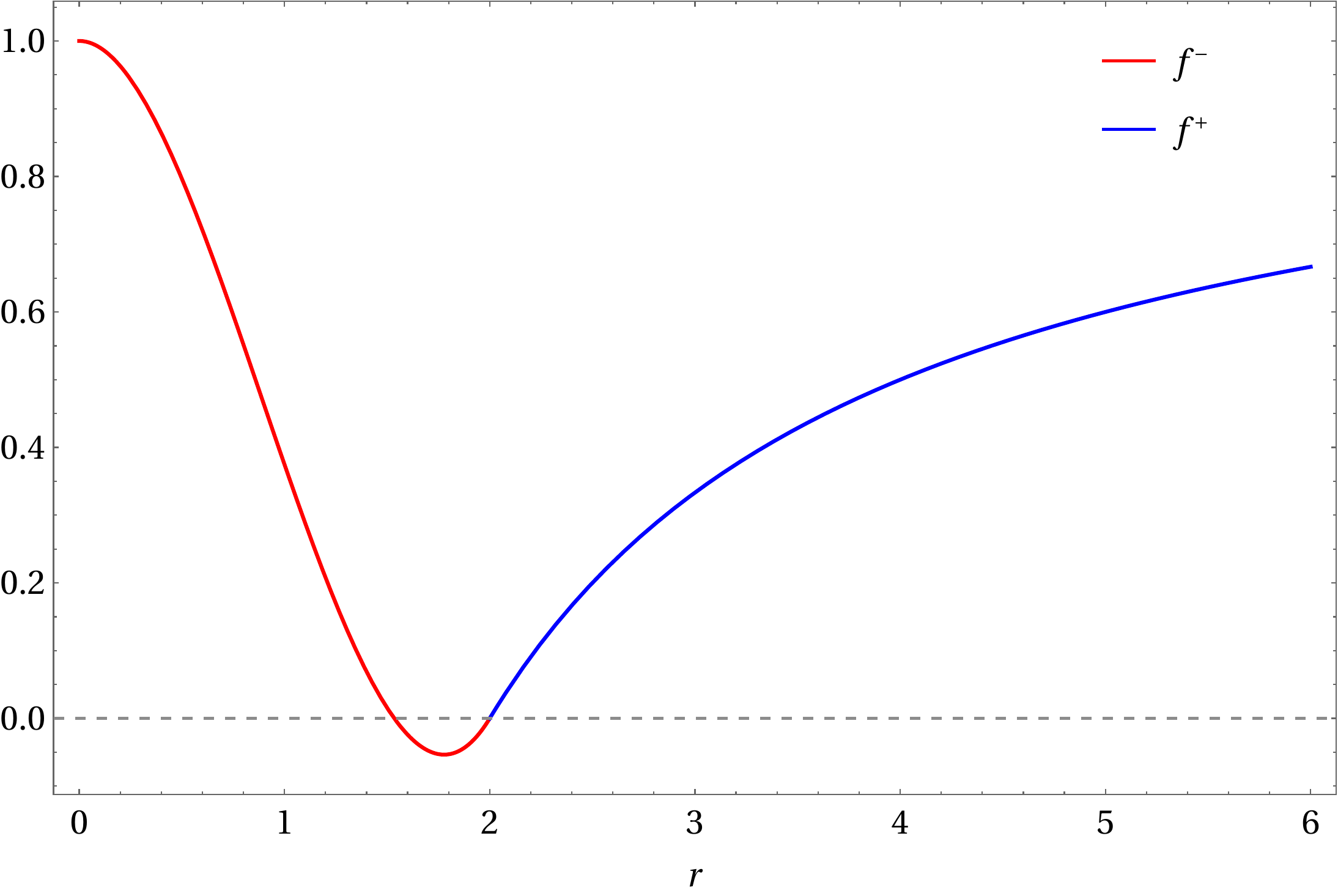}
$\ $
\\
\includegraphics[width=0.42\textwidth]{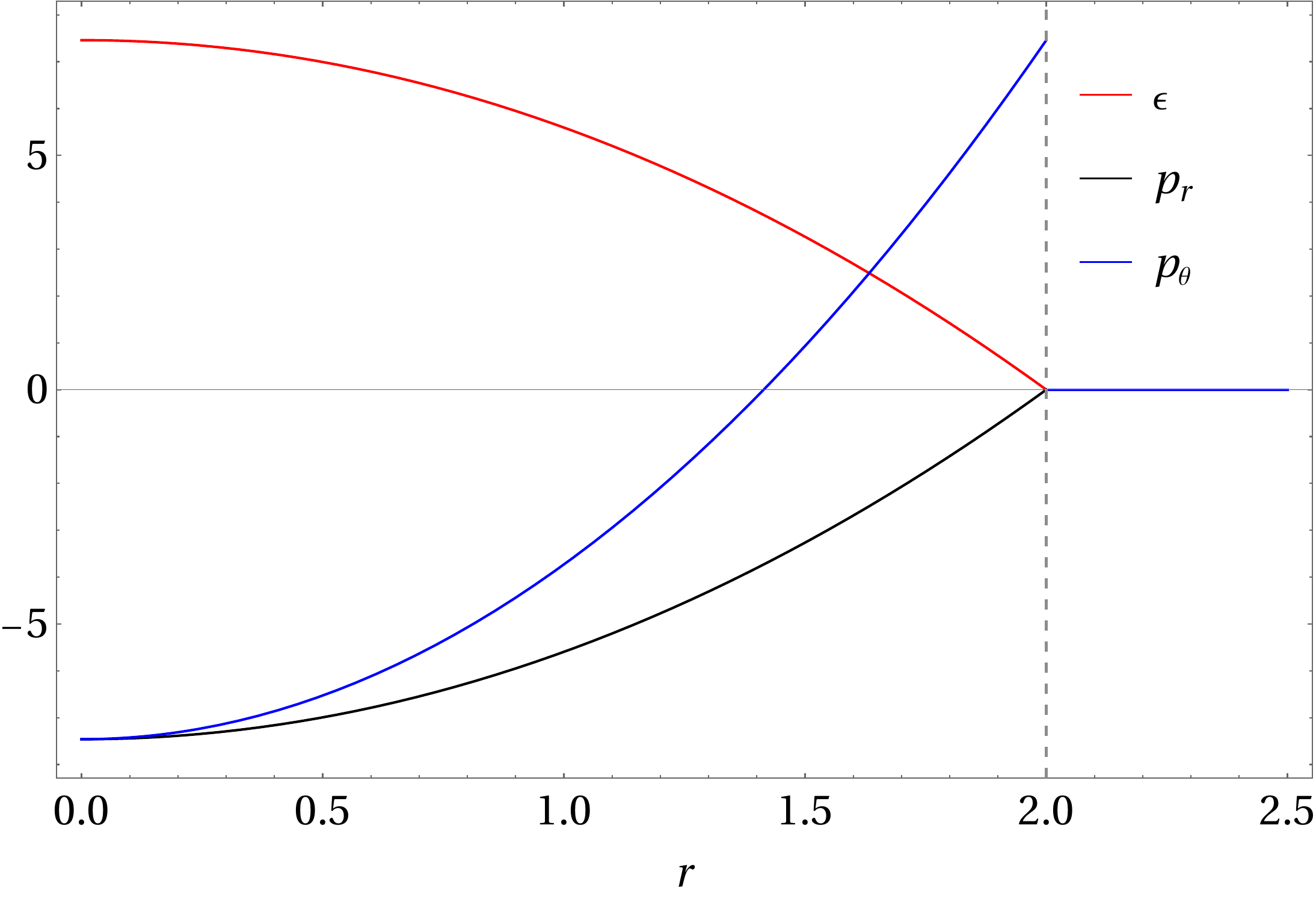}
\caption{Mass function~\eqref{m1} for $n=4$ (top panel),
corresponding metric function~\eqref{mtransform} (middle panel),
and density and pressures~\eqref{sources1a}-\eqref{sources1b}
(rescaled by a factor of $100$ for convenience; bottom panel).
Vertical dashed lines represent the horizon $h=2\,{\cal M}$.
The Cauchy horizon is at $h_{\rm c}=\sqrt{2/3}\,h$.
(All quantities in units of ${\cal M}$.)}
\label{fig1}
\end{figure}

The simplest regular solution with Schwarzschild exterior was found by imposing
the continuity conditions~\eqref{cond2} on the mass function~\eqref{M}
[see Ref.~\cite{Ovalle:2024wtv} for all details], which yields
\begin{equation}
\label{m1}
m
=
\frac{r}{2\,(n-2)}
\left[
\left(n+1\right)\frac{r^2}{h^2}
-3\left(\frac{r}{h}\right)^{n}
\right]
\ ,
\end{equation}
where $2<n\in\mathbb{N}$ is a parameter (not hair) which labels a family of
regular BHs, and whose physical interpretation will be elucidated later.
The corresponding metric function reads
\begin{equation}
\label{sol1}
f^-
=
1
-\frac{1}{n-2}
\left[
n+1
-3\left(\frac{r}{h}\right)^{n-2}
\right]
\frac{r^2}{h^2}
\ ,
\quad
\end{equation}
which gives rise to the curvature
\begin{equation}
\label{Rsin1}
R
=
\frac{n+1}{n-2}
\left[4-\left(n+2\right)\left(\frac{r}{h}\right)^{n-2}\right]
\frac{3}{h^2}
\ ,
\end{equation}
and is sourced by a fluid with
\begin{equation}
\label{sources1a}
\kappa\,\epsilon
=
-\kappa\,p_r
=
\frac{n+1}{n-2}
\left[1-\left(\frac{r}{h}\right)^{n-2}\right]
\frac{3}{h^2}
\end{equation}
and
\begin{equation}
\kappa\,p_\theta
=
\frac{n+1}{n-2}
\left[\frac{n}{2}
\left(\frac{r}{h}\right)^{n-2}-1\right]
\frac{3}{h^2}
\ ,
\label{sources1b}
\end{equation} 
with all expressions of course valid for $0<r\le h$.
These fluids sourcing the BH solutions~\eqref{sol1} satisfy the weak energy condition,
and represent alternative sources for the Schwarzschild exterior $r>h$ in
Eq.~\eqref{mtransform}.
Moreover, density and pressures behave monotonically, as we can see for $n=4$
in Fig.~\ref{fig1}.
\begin{figure}
\includegraphics[width=0.42\textwidth]{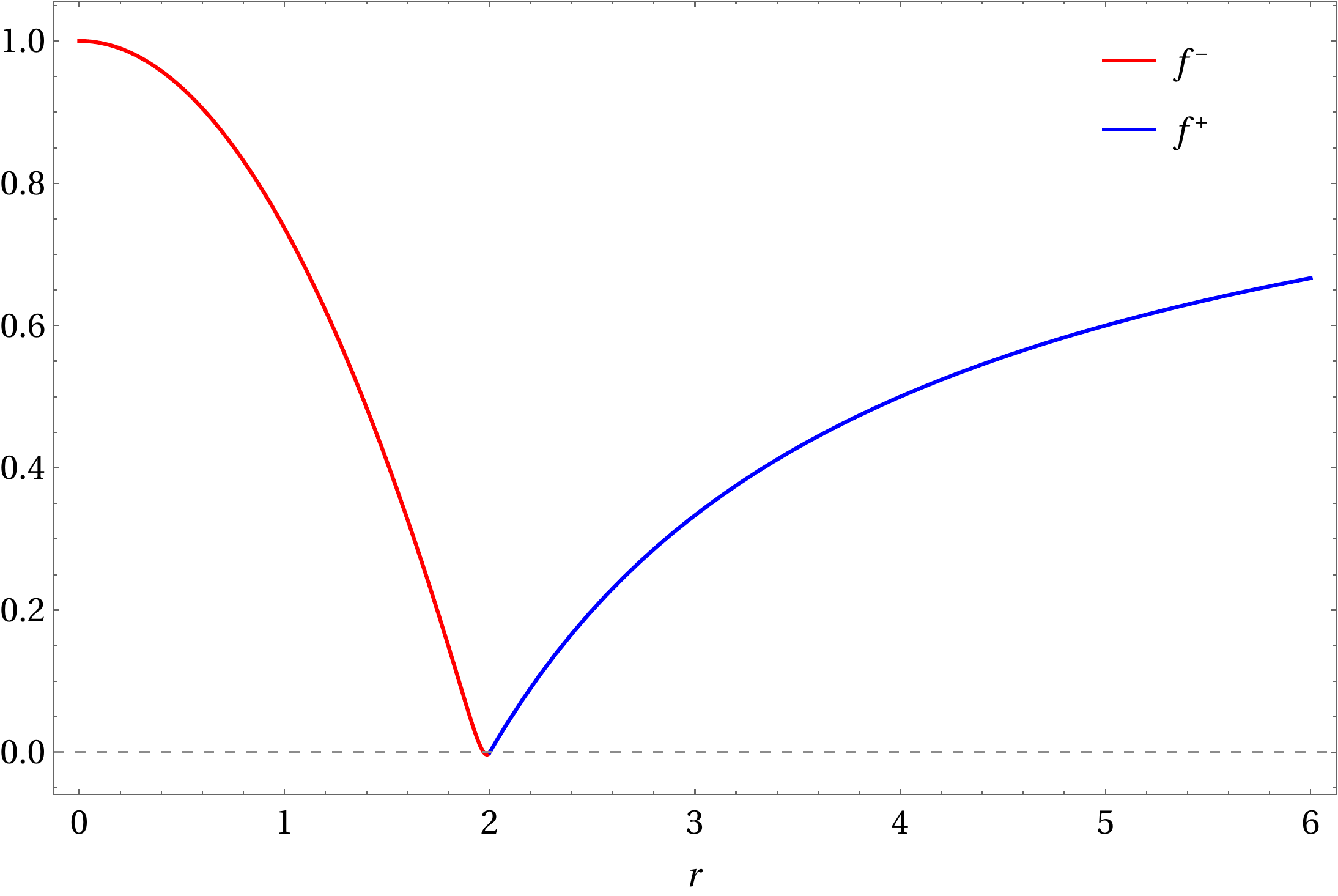}
\\
$\ $
\\
\includegraphics[width=0.42\textwidth]{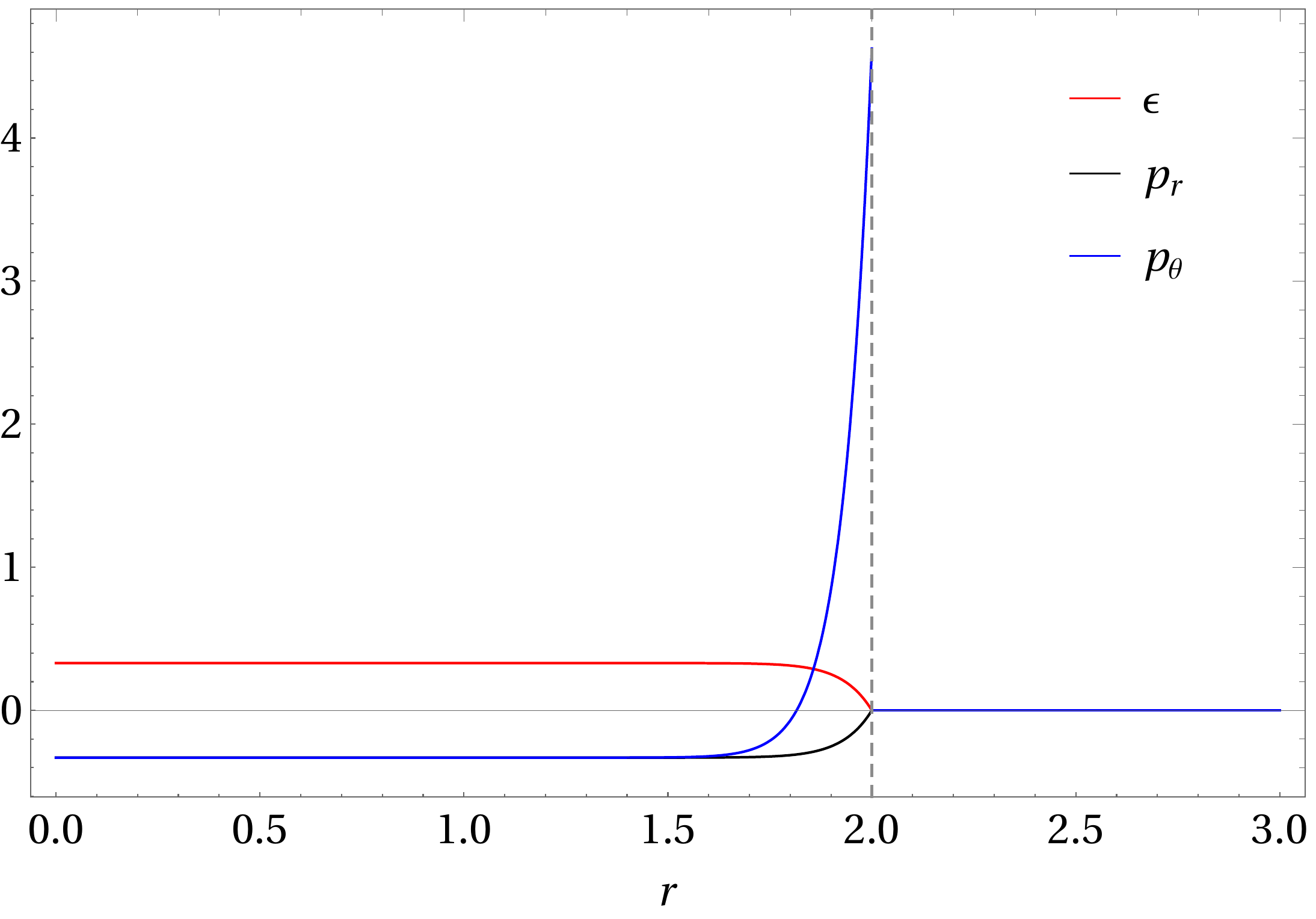}
\caption{Metric function~\eqref{mtransform} for $n=30$
(upper panel), and corresponding density and
pressures~\eqref{sources1a}-\eqref{sources1b} (rescaled by a
factor of $10$ for convenience; lower panel).
Vertical dashed line represents the horizon $h=2\,{\cal M}$ coincident
with the Cauchy horizon.
(All quantities in units of ${\cal M}$.)}
\label{fig2}
\end{figure}
\begin{figure}
\includegraphics[width=0.42\textwidth]{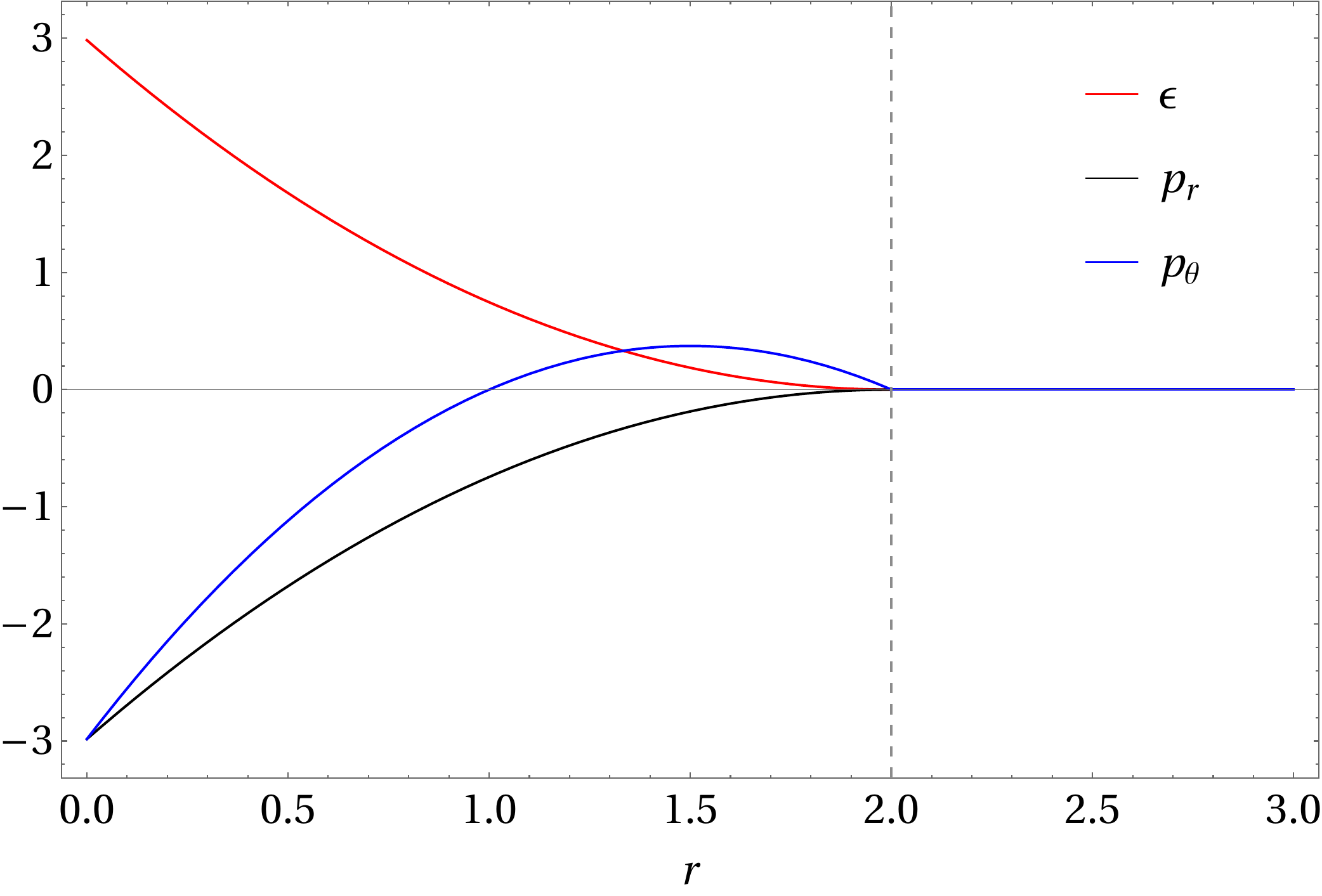}
\\
$\ $
\\
\includegraphics[width=0.42\textwidth]{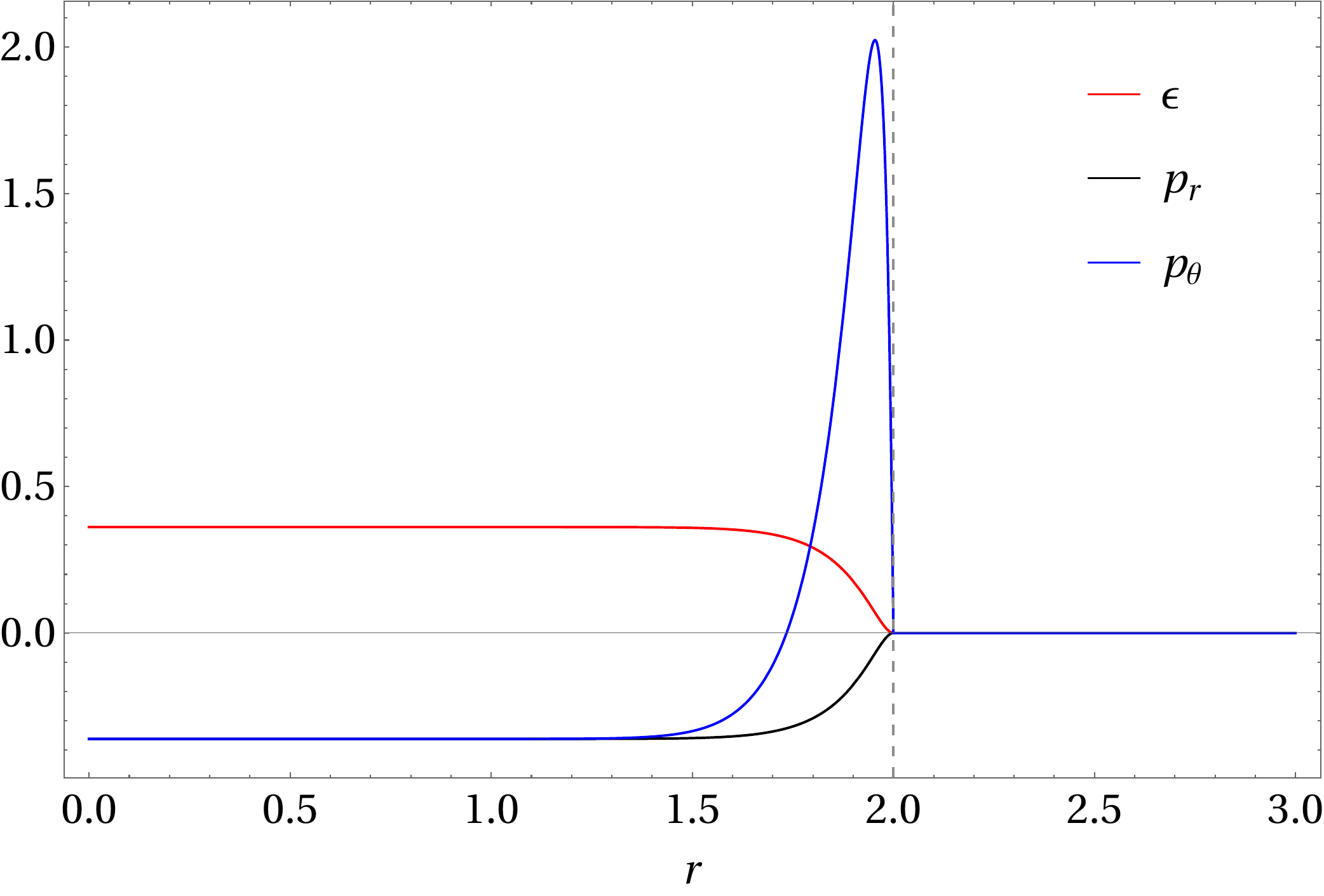}
\caption{Density and pressures (rescaled by a factor of 10 for convenience)
for the solution~\eqref{sol2} with $n=3$ and $l=4$ corresponding to the mass
function~\eqref{m34} (upper panel), and the quasi-extremal case for $l=4\,n=80$
with the Cauchy and event horizon almost merging at $r=h$ (lower panel).
Vertical dashed lines represent the horizon $h=2\,{\cal M}$.
(All quantities in units of ${\cal M}$.)}
\label{fig3}
\end{figure}
\par
We further notice that the region near $r=0$ always behaves like the de~Sitter solution
with effective cosmological constant $\Lambda_{eff}=3/h^2$, namely
\begin{equation}
\label{dS}
f
\sim
1-{r^2}/{h^2}
\ ,
\end{equation}
with
\begin{equation}
\kappa\,\epsilon
=
-\kappa\,p_r
\sim
-\kappa\,p_\theta
\sim
{3}/{h^2}
\sim
{R}/{4}
\ .
\end{equation}
For increasing $n\gg 2$, this de~Sitter core grows towards the horizon,
where the behaviour changes in order to match the Schwarzschild exterior,
as illustrated for $n=30$ in Fig.~\ref{fig2}.
\subsection{Regular Schwarzschild BHs with $p_\theta(h)=0$}
\label{SS:rrBH}
We have seen that, unlike the density and radial pressure, the tension $p_\theta$
does not need to vanish on the horizon when the metric in the exterior is given
by the Schwarzschild geometry.
In fact, Eq.~\eqref{sources1b} yields a finite value $p_\theta\propto h^{-2}$ at $r=h$. 
\par
A family of regular solutions with smoother match with the outer
Schwarzschild metric can be found by further imposing
\begin{equation}
\label{cond3}
p_\theta(h)
\propto
m''(h)
=
0
\ ,
\end{equation}
which yields
\begin{eqnarray}
\label{Mr}
m
&=&
\frac{(n+1)(l+1)}{(n-2)(l-2)}
\left[\left(\frac{r}{h}\right)^2
-\frac{3\,(l-2)}{(n+1)(l-n)}\left(\frac{r}{h}\right)^n
\right.
\nonumber
\\
&&
\left.
+\frac{3\,(n-2)}{(l+1)(l-n)}\left(\frac{r}{h}\right)^l \right]
\frac{r}{2}
\ ,
\end{eqnarray}
with $2<n<l\in\mathbb{N}$.
(Notice that Eq.~\eqref{Mr} is invariant under the exchange $l\leftrightarrow n$.)
The corresponding metric function reads
\begin{eqnarray}
\label{sol2}
f^-
&=&
1-
\frac{(n+1)(l+1)}{(n-2)(l-2)}
\left[\left(\frac{r}{h}\right)^2
-\frac{3\,(l-2)}{(n+1)(l-n)}\left(\frac{r}{h}\right)^n
\right.
\nonumber
\\
&&
\phantom{A\,B\,\frac{(n+1)(l+1)}{(n-2)(l-2)}}
\left.
+\frac{3\,(n-2)}{(l+1)(l-n)}\left(\frac{r}{h}\right)^l \right]
\ ,
\end{eqnarray}
and the source is characterised by
\begin{eqnarray}
\label{sources2a}
\kappa\,\epsilon
&=&
\left[\left(\frac{r}{h}\right)^2
+\frac{n-2}{l-n}\left(\frac{r}{h}\right)^{l}
-\frac{l-2}{l-n}\left(\frac{r}{h}\right)^{n}\right]
\frac{3}{r^2}
\nonumber
\\
&&
\phantom{\,}
\times\frac{\left(n+1\right)\left(l+1\right)}{\left(n-2\right)\left(l-2\right)}
=
-\kappa\,p_r
\end{eqnarray} 
and
\begin{eqnarray}
\kappa\,p_\theta
&=&
-
\left[\left(\frac{r}{h}\right)^2
+
\frac{l\left(n-2\right)}{2\left(l-n\right)}\left(\frac{r}{h}\right)^{l}
-\frac{n\left(l-2\right)}{2\left(l-n\right)}\left(\frac{r}{h}\right)^{n}\right]
\frac{3}{r^2}
\nonumber
\\
&&
\phantom{A}
\times
\frac{\left(n+1\right)\left(l+1\right)}{\left(n-2\right)\left(l-2\right)}
\ .
\label{sources2b}
\end{eqnarray} 
The above solution represents the simplest regular BH with Schwarzschild
exterior having continuous energy-momentum tensor across the horizon,
{\em i.e.}~$T^\mu_{\,\,\nu}(h)=0$.
\par
Like the cases in Section~\ref{SS:rBH}, the solutions~\eqref{sol2} satisfy
the weak energy condition, and represent an alternative source for the
Schwarzschild exterior $r>h$ in Eq.~\eqref{mtransform}.
Moreover, the expressions~\eqref{Mr}-\eqref{sources2b} converge to
those in Eqs.~\eqref{m1}-\eqref{sources1b} for $l\gg n$ in the region
$r<h$. 
\par
Eq.~\eqref{Mr} for $l=4$ and $n=3$ reads
\begin{equation}
\label{m34}
m
=
\frac{r^3}{2\,h^4}\left(10\,h^2-15\,h\,r+6\,r^2\right)
\ .
\end{equation}
This is the same mass function for the metric found in semi-analytic form
in Ref.~\cite{Mars:1996khm}, which is however not of the Kerr-Schild
type.~\footnote{To our knowledge, this was the first work about a BH
with exact Schwarzschild exterior and distributed source.}
For a Kerr-Schild metric~\eqref{metric}, the mass function~\eqref{m34}
results in a Cauchy inner horizon at $r=h/2$ and the source
is characterised by
\begin{equation}
\label{sources22a}
\kappa\,\epsilon
=
-\kappa\,p_r
=
\frac{2}{r^2\,h^3}
\left(h-r\right)^2
\left(h+2\,r\right)
\end{equation}
and
\begin{equation}
\kappa\,p_\theta
=
\frac{6}{h^3}
\left(h-r\right)
\ ,
\label{sources22b}
\end{equation}
with curvature 
\begin{equation}
\label{Rsin2}
R
=
\frac{4}{r^2}\left(1+\frac{5\,r^3}{h^3}-\frac{6\,r^2}{h^2}\right)
\ ,
\end{equation}
in the interior $0< r\leq h$.
\par
\subsection{Extremal Schwarzschild BHs}
\label{SS:eBH}
\begin{table*}
\caption{Interior of regular Schwarzschild BHs with mass functions~\eqref{m1} and~\eqref{Mr}.
\label{tab1}}
\begin{ruledtabular}
\begin{tabular}{ c c c c }
	$\{n,\,l\}$  & $m(r)$ &$\epsilon>0$& Energy condition
	\\
	\hline\hline
	$\{2<n<l\}$
	&
	$m
	=
	\frac{(n+1)(l+1)}{(n-2)(l-2)}\left[\left(\frac{r}{h}\right)^2-\frac{3\,(l-2)}{(n+1)(l-n)}\left(\frac{r}{h}\right)^n
	+\frac{3\,(n-2)}{(l+1)(l-n)}\left(\frac{r}{h}\right)^l \right]\frac{r}{2}$  &Yes &Weak
	\\
	\hline
	$\{2<n\ll l\}$
	&
	$m
	\sim
	\frac{r}{2(n-2)}\left[\frac{r^2}{h^2}\left(n+1\right)-3\left(\frac{r}{h}\right)^n\right]$  &Yes &Weak
	\\
	\hline
	$\{2\ll n\ll l\}$
	&
	$m
	\sim
	\frac{r^3}{2h^2}\quad$ Extremal BH with $h_{\rm c}\sim\,h$  &Yes &Weak
	\\ 
	\hline
	$\{2<n,l\to \infty\}$ & 
	$m
	=
	\frac{r}{2(n-2)}\left[\frac{r^2}{h^2}\left(n+1\right)-3\left(\frac{r}{h}\right)^n\right]$
	&Yes & Weak
	\\ 
	\hline
	$\{2\ll n,l\to \infty\}$ & 
	$m
	\sim
	\frac{r^3}{2h^2}\quad $ Extremal BH with $h_{\rm c}\sim\,h$&Yes& Weak
	\\  
	\end{tabular}
	\end{ruledtabular}
\end{table*}
It is easy to see that all of the regular BH interiors described by the metric 
functions~\eqref{sol1} and~\eqref{sol2} of the previous Sections have a single
inner horizon.
A rather interesting aspect of these solutions is then that they also contain
configurations that are almost extreme black holes, as we will see below.
\par
The interior metric function $f^-$ has a minimum,
\begin{equation}
\label{extrema}
(f^-)'(\re)
=
0
\ ,
\end{equation}
which, for the solutions~\eqref{sol1}, is located at
\begin{equation}
\label{extrema2}
\re
=
h\left[\frac{2}{3}\left(1+\frac{1}{n}\right)\right]^{\frac{1}{n-2}}
\ .
\end{equation}
Since 
\begin{equation}
	\label{hc-to-h}
\lim_{n\to\infty} \re
=
h
\ ,
\end{equation}
this minimum shifts towards the event horizon for increasing values 
of $n$.
Consequently, the Cauchy horizon given by $f^-(h_{\rm c})=0$, with
$h_{\rm c}<h$, is located at
\begin{equation}
h_{\rm c}
\sim
h
\ ,
\end{equation}
for $n\gg 2$.
Such configurations represent quasi-extremal BHs with de Sitter core
and Schwarzschild exterior separated by a (infinitesimally thin) region
$h_{\rm c}<r<h$, as displayed in  Figure~\ref{fig2}.
The parameter $n$ for the solutions~\eqref{sol1} therefore measures
how close to extremality the object is.
\par
Since the solutions~\eqref{sol1} are a particular case of the expression in
Eq.~\eqref{sol2} for $l\gg n$, we conclude that the solutions~\eqref{sol2}
must have the same causal structure.
It is still important to remark that the solutions~\eqref{sol2} are generated
by fluids whose energy-momentum tensor is completely continuous
across the event horizon $r=h$, as displayed in Fig.~\ref{fig3}. 
The solutions analysed in Section~\ref{SS:rrBH} can therefore be viewed as an
improvement over those in Section~\ref{SS:rBH}.
\par
We can obtain even smoother solutions by considering a generic polynomial
of $N$ terms of the form
\begin{equation}
	\label{mpoly}
	m
	=
	C_3\,r^3
	+
	C_n\,r^n
	+
	C_l\,r^l
		+
	C_p\,r^p
	+
	\ldots
	\ ,
\end{equation}
where the unknown coefficients can be determined by the continuity
condition~\eqref{abcd} and
\begin{equation}
\label{cond-n}
\frac{d^q m}{dr^q}(h)
=
0
\ ,
\end{equation}
for all $1\le q\leq N-1$.
As we have seen, full continuity of the energy-momentum tensor across
the horizon is obtained for $N=3$, corresponding to the inner metric
functions~\eqref{sol2}.
A summary of the cases with $N=3$ and $N=2$ is given in Table~\ref{tab1}
for convenience. However, as we have just seen, the inner regular region 
could be much richer than illustrated in Table~\ref{tab1}.
\section{Cosmology}
\label{sec4}
For all the regular BHs described in Table~\ref{tab1}, the region $h_{\rm c}<r< h$ lies
between two horizons and can also be considered as a whole universe~\cite{Doran:2006dq}.
{In fact, the metric signature is $(+,-,+,+)$ inside this region where
$r$ becomes a time coordinate.
To make the role of time and spatial coordinates more explicit, we can swap
$t\leftrightarrow r$ therein, so that the corresponding line element reads
\begin{eqnarray}
\label{patch2}
ds^{2}
=
-\frac{dt^2}{F(t)}
+
F(t)\,dr^{2}
+t^2\,d\Omega^2
\ ,
\end{eqnarray}
where
\begin{equation}
\label{F}
F
=
1-\frac{2\,m(t)}{t}
\geq
0
\end{equation}
with the mass function $m$ given by the different cases listed in Table~\ref{tab1} and $h_{\rm c}\equiv t_1<t<t_0\equiv h$.}
\par
We can next write the metric~\eqref{patch2} in terms of the cosmic 
(or synchronous) time defined by
\begin{equation}
\label{tau}
d\tau
=
\pm\frac{dt}{\sqrt{F(t)}}
\ ,
\end{equation}
which leads to the generic cosmological solution
\begin{equation}
\label{cosmo}
ds^2
=
-d\tau^2
+a^2(\tau)\,dr^{2}
+b^2(\tau)\,d\Omega^2
\ .
\end{equation}
The metric~\eqref{cosmo} represents a Kantowski-Sachs homogeneous
but anisotropic universe~\cite{Kantowski:1966te,Brehme:1977fi}
with scale factors
 \begin{eqnarray}
\label{scalefactors}
a^2(\tau)
&\equiv&
F(\tau)
\nonumber
\\ 
\\
b^2(\tau)
&\equiv&
t^2(\tau)
\ .
\nonumber
\end{eqnarray}
The non-vanishing components of the corresponding Einstein tensor 
are given by
\begin{eqnarray}
\label{einsteintensor}
G^0_{\ 0}
&=&
-\left(\frac{1}{b^2}
+\frac{2\,\dot{a}\,\dot{b}}{a\,b}
+\frac{\dot{b}^2}{b^2}\right)
\\ 
G^1_{\ 1}
&=&
-\left(\frac{1}{b^2}+\frac{2\,\ddot{b}}{b}+\frac{\dot{b}^2}{b^2}\right)
\\ 
G^2_{\ 2}
&=&
-\left(\frac{\dot{a}\,\dot{b}}{a\,b}+\frac{\ddot{a}}{a}+\frac{\ddot{b}}{b}\right)
\ ,
\end{eqnarray}
where dots denote derivative with respect to $\tau$.
\par
Let us then consider a regular BH described in the first line of Table~\ref{tab1},
with $n=3$ and $l=4$, so that the function in Eq.~\eqref{F} reads
\begin{equation}
F
=
6\left(\frac{t}{h}-1\right)
\left(\frac{t}{h}-\frac12\right)
\left(\frac{t^2}{h^2}-\frac{t}{h}-\frac13\right)
\ ,
\label{F34}
\end{equation}
where $t$ runs between the two horizons at $t = h$ and $t=h/2$.
The corresponding expression for the cosmic time~\eqref{tau} is integrable,
leading to a finite lapse between $t=h/2$ and $t=h$.
If we choose the initial value $\tau = 0$ corresponding to $t=h/2$, the
final time can be computed numerically and is given by
\begin{equation}
\tau_0
\equiv
\tau(h)
=
\int_{1/2}^{1}\frac{d t}{\sqrt{F(t)}}
\simeq
1.9 \,h
\ .
\label{final}
\end{equation}  
\par
In the vicinity of the point $t = {h}/{2}$, we can write 
\begin{equation}
t
=
\frac{h}{2}
+
y
\ ,
\label{y}
\end{equation} 
with $0\le y\ll h/2$,
and we find
\begin{equation}
\tau
\simeq
2\,\sqrt{\frac{h}{7}}
\int_{\frac{h}{2}}^{\frac{h}{2} + y}
\frac{d x}{\sqrt{x}}
=
4\,\sqrt{\frac{h\,y}{7}}
\ .
\label{cosmic1}
\end{equation}
Correspondingly, the components of the Kantowski-Sachs metric read
\begin{equation}
b^2
=
t^2
\simeq
\frac{h^2}{4}
\left(
1
+
\frac{7\,\tau^2}{4\,h}
\right)
\ ,
\label{time}
\end{equation}
and 
\begin{equation}
a^2
=
F
\simeq
\frac{49\,\tau^2}{64\,h^2}
\ .
\label{A1}
\end{equation}
One can directly check that the scalar curvature
\begin{equation}
R=
-\frac{\ddot{a}}{a}
-4\,\frac{\ddot{b}}{b}
-4\,\frac{\dot{a}\,\dot{b}}{a\,b}
-2\,\frac{\dot{b}^2}{b^2}
-\frac{2}{b^2}
\label{curv}
\end{equation}
is regular near $\tau=0$.
Other invariants are also not singular and the universe is regular at $\tau = 0$. 
\par
We can repeat the analysis near $t = h$, by defining
\begin{equation}
t = h-y
\ ,
\label{time1}
\end{equation}
where again $0\le y\ll h/2$.
The cosmic time then reads
\begin{equation}
\tau
\simeq
\tau_0
-2\,\sqrt{h\,y}
\ ,
\label{tau2}
\end{equation}
and the Kantowski-Sachs metric functions are given by 
\begin{equation}
b^2
=
t^2
\simeq
h^2
\left[
1
-
\frac{(\tau_0-\tau)^2}{2\,h^2}
\right]
\end{equation}
and
\begin{equation}
a^2
=
F
\simeq
\frac{(\tau_0-\tau)^2}{4\,h^2}
\ .
\label{a1}
\end{equation}
This point is also non-singular. 
\par
Overall, the above case corresponds to a quasi-cyclic universe
which starts at $\tau=0$ with $a=0$ and $b=h/2$ (the radius
of the inner horizon).
As the synchronous time increases, both $a$ and $b$ increase,
but $a$ reaches a maximum at $\tau=\tau_1$ determined by the
condition
\begin{equation}
\frac{d F}{d t}
=
\frac{t}{h}
\left(
24\,\frac{t^2}{h^2}
-45\,\frac{t}{h}
+20
\right)
=
0
\ ,
\label{t1}
\end{equation}
after which $a$ decreases and reaches zero at $\tau=\tau_0$, when
$b=h$ (the radius of the outer horizon).
It is easy to compute the solution of Eq.~\eqref{t1}, which is given  by
\begin{equation}
t_1
=
\frac{45-\sqrt{105}}{48}\,h
\simeq
0.72\,h
\ ,
\end{equation}
and the corresponding cosmic time can then be computed numerically,
to wit
\begin{equation}
\tau_1
\simeq
0.79\,h
\ .
\end{equation}
\par
Obviously, the first derivative of the scale factor $b$ is always positive,
since
\begin{equation}
\dot{b}
=
\dot t
=
\sqrt{F}
=
a
\ .
\label{b-der}
\end{equation} 
Since $\ddot b=\dot a$,
when the scale factor $a$ is growing (for $0<\tau< \tau_1$), the scale factor 
$b$ grows with positive acceleration.
For $\tau_1<\tau<\tau_0$, the scale factor $a$ is decreasing and
the scale factor $b$ increases with negative acceleration.
\par
This behaviour remains qualitative the same in all cases containing two horizons,
the main difference being the location (or time) of the inner horizon (the outer
horizon is fixed at $t=h$).
For example, for $n=3$ and $l=0$, one finds
\begin{equation}
F
=
3\left(1-\frac{t}{h}\right)
\left(\frac{t}{h}-\frac{1+\sqrt{13}}{6}\right)
\left(\frac{t}{h}-\frac{1-\sqrt{13}}{6}\right)
\ ,
\end{equation} 
and the inner horizon is at
\begin{equation}
t
=
\frac{1+\sqrt{13}}{6}\,h
\simeq
0.77\,h
\ ,
\end{equation} 
which is larger than the value $t=h/2$ for $n=3$ and $l=4$.
\par
\begin{figure}[t]
\centering
\includegraphics[width=0.45\textwidth]{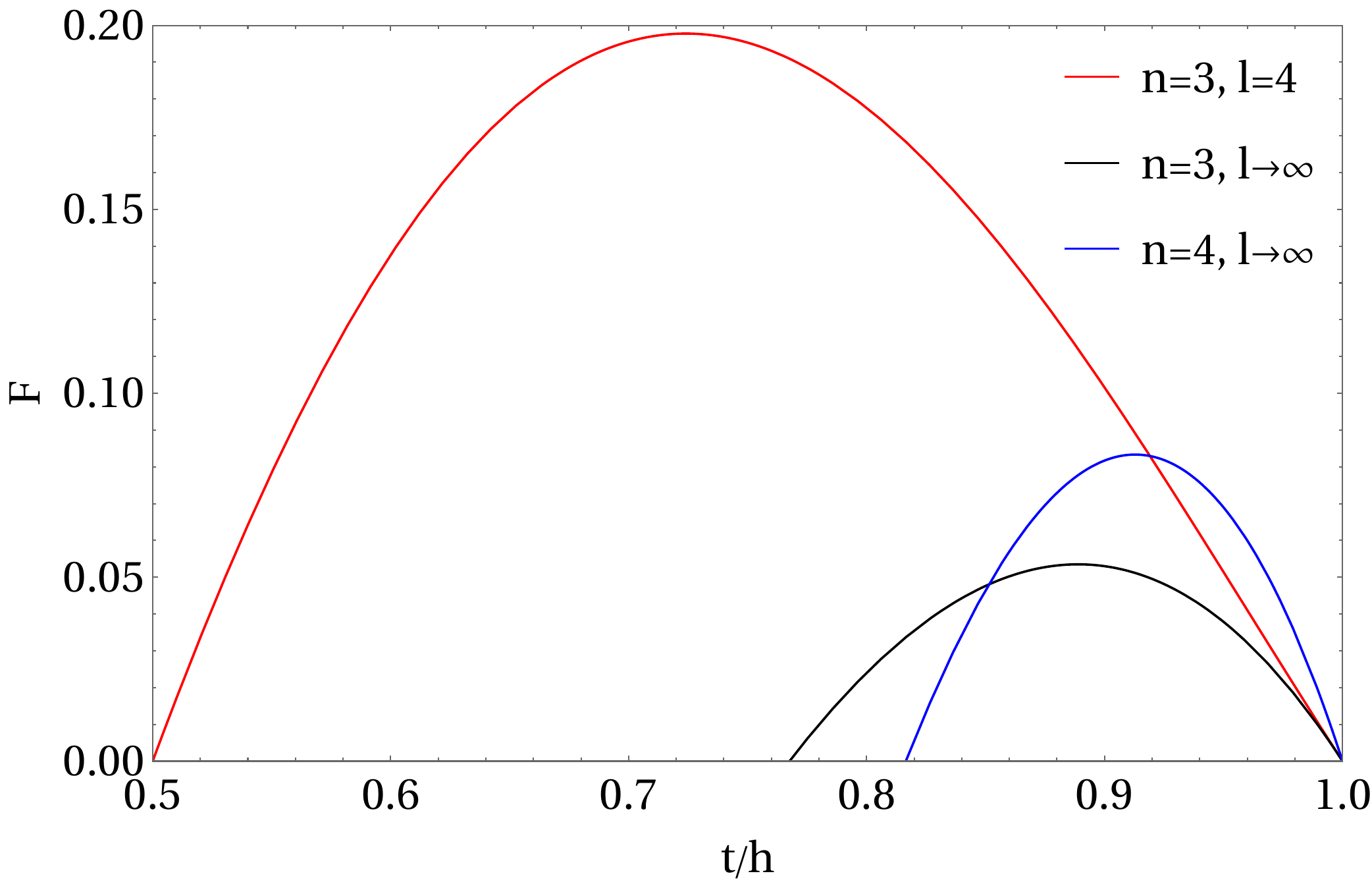}
\caption{Function $F$ in Eq.~\eqref{F} for the cases discussed in the
main text.}
\label{fig4}
\end{figure}

A particularly simple case is obtained for $n=4$ and $l=0$,
namely
\begin{equation}
F
=
3\left(1-\frac{t^2}{h^2}\right)
\left(\frac{t^2}{h^2}-\frac{2}{3}\right)
\ ,
\end{equation} 
in which the coordinate time runs in the interval
\begin{equation}
0.82\,h
\simeq
\sqrt{\frac{2}{3}}\,h
\le
t
\le
h
\ .
\end{equation} 
The function $F$ for the three cases considered here is shown
in Fig.~\ref{fig4}.
We finally remark that the sign of the proper time
$\tau$ can be inverted, according to the definition~\eqref{tau},
so that the curves in Fig.~\ref{fig4} could be followed in either
direction.
\section{Conclusion}
\label{sec5}
Circumventing the singularity theorem in GR
necessarily implies the existence of non-classical states of matter that somehow
manage to stop the collapse.
Regular configurations with finite energy density everywhere then always contain
an additional inner horizon.
Achieving this without introducing exotic matter and maintaining the Schwarzschild
exterior is challenging.
In this work, we obtained regular BHs sourced by a fluid that satisfies the
weak energy condition and are characterized by a single charge, namely,
the ADM mass ${\cal M}$ of the Schwarzschild exterior.
Moreover, {unlike the singular metrics considered previously in Ref.~\cite{Casadio:2024fol,Aoki:2024dyr}},
these regular BHs admit (quasi) extremal configurations in which the two horizon (almost) coincide.
At first glance, this might seem odd, since extremal configurations are
usually achieved by specific combinations of at least two charges (e.g.~electric charge
$Q={\cal M}$ for the Reissner–Nordstr\"om and angular momentum $a={\cal M}$
for the Kerr solutions).
\par
We remark that {the solutions considered in this work technically admit quasi-extremal}
configurations, since the two horizons are separated by a layer which can be made as thin as we want,
thus effectively merging them in the limit of Eq.~\eqref{hc-to-h}.
In the same limit, the surface gravity $\kappa(r)=F'(r)$ on the Cauchy horizon 
also becomes as small as possible, namely
\begin{equation}
\lim_{n\to\infty}
\kappa(h_{\rm c})
=
\lim_{n\to\infty}
F'(h_{\rm c})
=
0
\ .
\end{equation}
This fact is quite significant, and could have consequences in favour of the existence
of stable regular BHs, since the instability caused by the mass inflation is precisely
proportional to $\kappa$.
This is a point that certainly deserves to be investigated further, since the existence of
extremal BHs described only by the ADM mass ${\cal M}$, which also happen
to be stable, would greatly support GR as the theory that correctly describes
very compact objects.
\par
We have also studied non-extremal configurations and showed that the layer between
the two horizons describes anisotropic Kantowski-Sachs universes, which show a
quasi-periodic evolution and contain no singularity at the end-points.
\subsection*{Acknowledgments}
R.C.~and A.K.~are partially supported by the INFN grant FLAG.
The work of R.C.~has also been carried out in the framework of activities of the
National Group of Mathematical Physics (GNFM, INdAM).
J.O.~is partially supported by ANID FONDECYT~Grant No.~1210041.
%
%
%
%
\bibliography{references.bib}
\bibliographystyle{apsrev4-1.bst}
%
%
\end{document}